\documentclass[journal,twoside,web]{ieeecolor2}
\usepackage{generic}
\usepackage{cite}
\usepackage{amsmath,amssymb,amsfonts}
\usepackage{algorithmic}
\usepackage{graphicx}
\usepackage[hidelinks]{hyperref}
\usepackage{textcomp}
\usepackage{booktabs}
\def\BibTeX{{\rm B\kern-.05em{\sc i\kern-.025em b}\kern-.08em
    T\kern-.1667em\lower.7ex\hbox{E}\kern-.125emX}}
\begin{document}
\title{System Filter-Based Common Components Modeling for Cross-Subject EEG Decoding}
\author{Xiaoyuan Li, Xinru Xue, Bohan Zhang, Ye Sun, Shoushuo Xi and Gang Liu*
\thanks{This work was supported by the National Natural Science Foundation of China under Grant 62303423, in part by the STL 2030-Major Project under Grant 2022ZD0208500, in part by Postdoctoral Science Foundation of China under Grant 2024T170844 and Grant 2023M733245, in part by the Henan Province key resaerch and development and promotion of special projects under Grant 242102311239, in part by Shaanxi Provincial Key Research and Development Program under Grant 2023GXLH-012. (Corresponding author: Gang Liu. Email: gangliu\_@zzu.edu.cn)}
\thanks{X. Li, X. Xue, Y. Sun, S. Xi and G. Liu are with the college of Electrical and Information Engineering, Zhengzhou University, Zhengzhou, 450001, China (e-mail: lixiaoyuan@zzu.edu.cn; xue@gs.zzu.edu.cn; ye@gs.zzu.edu.cn; xss@gs.zzu.edu.cn; gangliu\_@zzu.edu.cn).}
\thanks{B. Zhang is with the School of Computer Science and Engineering, University of New South Wales, Sydney, NSW 2052, Australia (e-mail: z5465167@ad.unsw.edu.au).}}

\maketitle

\begin{abstract}
Brain–computer interface (BCI) technology enables direct communication between the brain and external devices through electroencephalography (EEG) signals. However, existing decoding models often mix common and personalized components, leading to interference from individual variability that limits cross-subject decoding performance. To address this issue, this paper proposes a system filter that extends the concept of signal filtering to the system level. The method expands a system into its spectral representation, selectively removes unnecessary components, and reconstructs the system from the retained target components, thereby achieving explicit system-level decomposition and filtering. We further integrate the system filter into a Cross-Subject Decoding framework based on the System Filter (CSD-SF) and evaluate it on the four-class motor imagery (MI) task of the BCIC IV 2a dataset. Personalized models are transformed into relation spectrums, and statistical testing across subjects is used to remove personalized components. The remaining stable relations, representing common components across subjects, are then used to construct a common model for cross-subject decoding. Experimental results show an average improvement of 3.28\% in decoding accuracy over baseline methods, demonstrating that the proposed system filter effectively isolates stable common components and enhances model robustness and generalizability in cross-subject EEG decoding.
\end{abstract}


\begin{figure*}[htbp]
    \centering
    \includegraphics[width=1.0\textwidth]{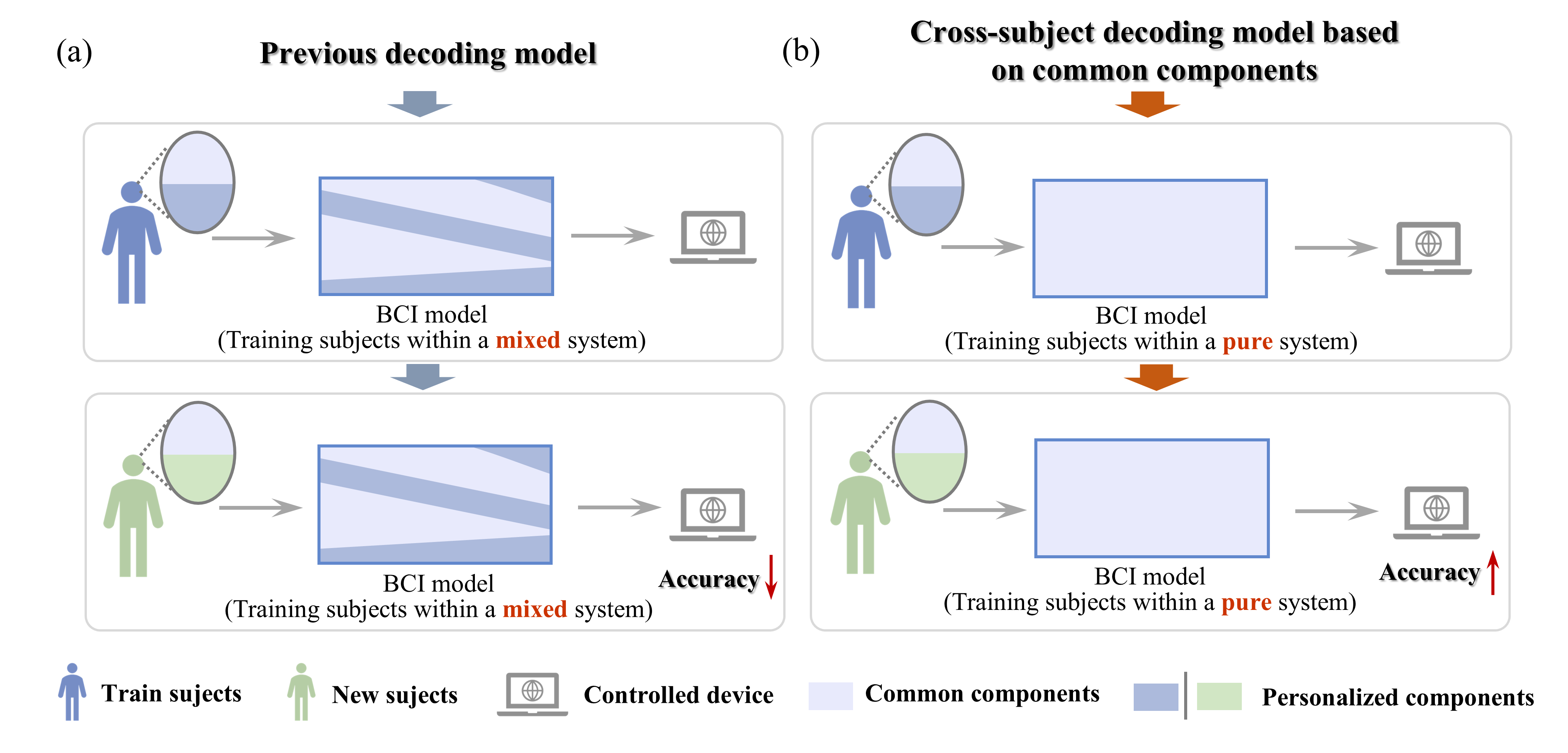}  
    \caption{The research goals of this paper. (a) Previous decoding models mix common and personalized components, leading to poor generalization for new subjects.
    (b) The proposed approach filters out personalized components and builds a pure cross-subject model based only on common components, enabling stable decoding.
    }
    \label{fig:li1}
\end{figure*}
\begin{figure}
    \centering
    \includegraphics[width=1.0\linewidth]{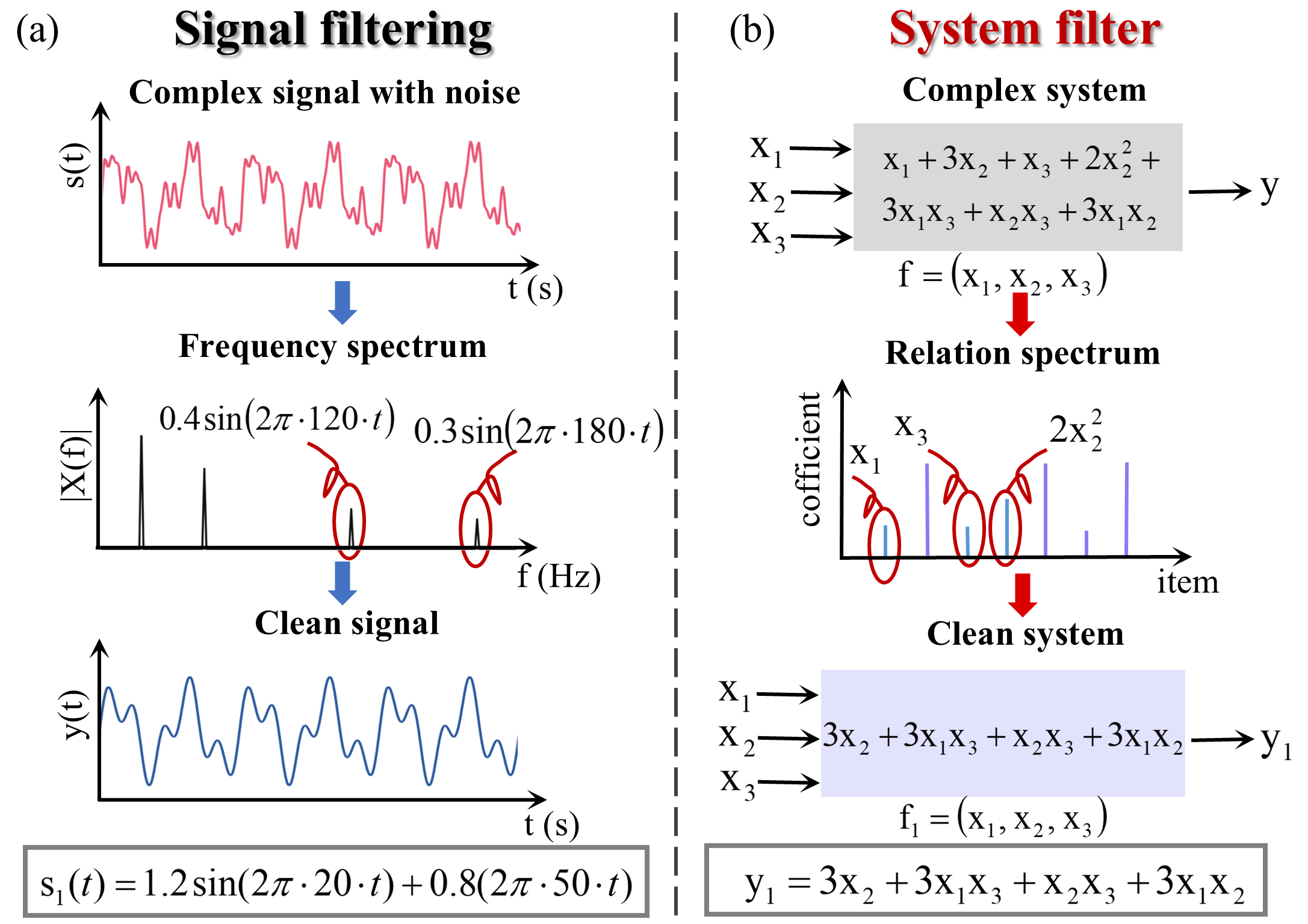}
    \caption{Conceptual analogy between the proposed system filter and signal filtering. (a) Signal filtering. A complex signal with noise is mapped into the frequency domain via Fourier transform and decomposed into multiple components. Filtering out unnecessary frequencies (e.g., 120 Hz, 180 Hz) yields a clean waveform containing only the target components (e.g., 20 Hz, 50 Hz). (b) System filter. A complex system, represented as a sum of polynomial terms, can be transformed into a spectral representation. By removing unnecessary terms (e.g., $x_1$, $x_3$, $x_i^2$) and retaining only the target ones, a purer and more interpretable system is obtained.
    }
    \label{fig:li2}
\end{figure}

\section{Introduction}
\label{sec:introduction}
\IEEEPARstart{B}{rain-computer} interface (BCI) systems decode electroencephalography (EEG) signals to directly control external devices, bypassing muscular and neural pathways. They have been widely applied in neurorehabilitation, motor assistance, and intelligent device control \cite{1.1,1.2,1.3,1.4}. However, cross-subject BCI decoding remains a significant challenge, as inter-subject variability substantially reduces the generalization capability of BCI models\cite{1.5,1.6,1.7}. This variability, which arises from subject-specific differences in brain activity patterns, makes it difficult for models to generalize effectively across different individuals\cite{111}. As a result, improving the robustness of BCI models and enhancing their generalization across subjects is a critical issue that needs to be addressed.

Conventional BCI decoding models are hybrid systems, combining two components: common components, which capture patterns shared across subjects, and personalized components, which capture subject-specific variations \cite{jeon2021mutual,gao2023double,chen2022transfer}. During training, these two elements are implicitly mixed into a single model. As illustrated in Fig.~\ref{fig:li1}(a), the purple region denotes common components, while the blue and green regions represent personalized components corresponding to different subjects. When a new subject is introduced, their EEG signals inevitably contain new personalized components, which were not present during training. These unseen components interfere with the trained hybrid model, leading to mismatches and reducing decoding accuracy \cite{1.8,zlatov2022towards}.

To address this problem, transfer learning (TL) has become the dominant approach\cite{scdan,zhang2021hybrid,tl,zanini2017transfer}. In a typical TL pipeline, a model is first pre-trained on data from multiple subjects and then fine-tuned with a small amount of data from a new subject. Fine-tuning emphasizes personalized features, adapting the model to the individual. However, this approach does not eliminate the interference caused by subject-specific components. While it improves within-subject performance, it does not solve the problem of cross-subject generalization, as the underlying hybrid structure remains sensitive to subject-specific variations.

This paper proposes a new approach that focuses solely on common components to construct a universal cross-subject decoding model, as shown in Fig.~\ref{fig:li1}(b). Rather than adapting a model to individual subjects by fine-tuning, the proposed method aims to filter out personalized components and retain only the common components shared across subjects. This approach seeks to eliminate the interference of subject-specific variability and improve the robustness and generalization of BCI decoding across different subjects. 

To achieve this goal, we propose a system filter, inspired by the principles of signal filtering, as illustrated in Fig.~\ref{fig:li2}. For example, Fourier transform maps a time-domain signal into the frequency domain, allowing for the selective removal of unnecessary frequency bands while retaining the essential signal\cite{Fouriertransform}. Similarly, the system filter transforms a complex system into a polynomial representation, maps it into spectral space, and selectively removes unnecessary components, finally reconstructs the system from the retained components. The human brain can be viewed as a biological system \cite{22}, where EEG recordings reflect brain activity that translates into specific intentions or motor commands. This EEG-based system serves as a biological information transmission channel, mapping neural activity to intended actions\cite{liu2022eegg}. By applying the system filter, we aim to isolate the stable, common patterns of brain activity across subjects, removing personalized variations. These retained common components are then used to construct a universal model, enabling more accurate and generalized cross-subject decoding.

The main contributions of this paper are as follows.

1 This paper proposes a system filter that transforms the system into spectrum representation, filters out unnecessary components, and retains only the target components, yielding a system composed solely of target components (see Fig.~\ref{fig:li2}). This method parallels signal filtering, where signals are decomposed in the frequency domain to remove noise.

2 This is the first study to apply the system filter to cross-subject motor imagery electroencephalography (MI-EEG) decoding tasks. By separating common components from subject-specific ones and constructing a universal model, experimental results demonstrate that the proposed system filter improves generalization, enhancing cross-subject decoding performance.

The remainder of this paper is organized as follows. Section 2 provides a detailed description of the system filter and its application in cross-subject MI-EEG decoding. Section 3 introduces experimental setups. Section 4 presents the experimental results. Section 5 discusses the implications of our findings. Section 6 concludes the paper.

\section{Methods}
\subsection{System Filter}
The proposed system filter follows a three-step procedure: expansion, filtering, and reconstruction, which are described in detail below.

\subsubsection{Expansion}
Any finite-dimensional system can be represented in the form of a polynomial expansion. 
Mathematically, based on the completeness of the polynomial basis, all internal relations of the system can be mapped onto a set of polynomial terms. 
Let $f : \mathbb{R}^d \to \mathbb{R}$ denote a system with input vector $x = (x_1, x_2, \ldots, x_d) \in \mathbb{R}^d$. 
To explicitly reveal its internal structure, the system is expanded into a relation spectrum.
\begin{equation}
    f(x) = \sum_{k=1}^{K} c_k \, \phi_k(x),
\end{equation}
where $\phi_k(x)$ denotes the $k$-th polynomial basis function, and $c_k$ is the corresponding coefficient. 
This expansion provides a complete and interpretable decomposition of the system into separable relation items.
\subsubsection{Filtering}
In the relation spectrum space, some relation items are necessary to the target output, while others correspond to unnecessary components. To filter out the unnecessary components, a selection operator $\mathcal{S}$ is defined as follows.
\begin{equation}
    \mathcal{S}(f) = \sum_{k \in \mathcal{I}} c_k \, \phi_k(x)
\end{equation}
where $\mathcal{I} \subseteq \{1, 2, \ldots, K\}$ denotes the index set of retained items. This operation performs “filtering” in the spectral space, yielding a purified representation of the target components.
\subsubsection{Reconstruction}
After filtering, the system is reconstructed by recombining the retained relation items.
\begin{equation}
    f^{*}(x) = \mathcal{S}(f)(x)
\end{equation}

This reconstruction yields a system composed solely of target components, eliminating the interference of unnecessary terms while achieving system-level decomposition and filtering in a compact, interpretable, and mathematically principled manner.

\begin{figure*}[t!]
    \centering
    \includegraphics[width=1.0\textwidth]{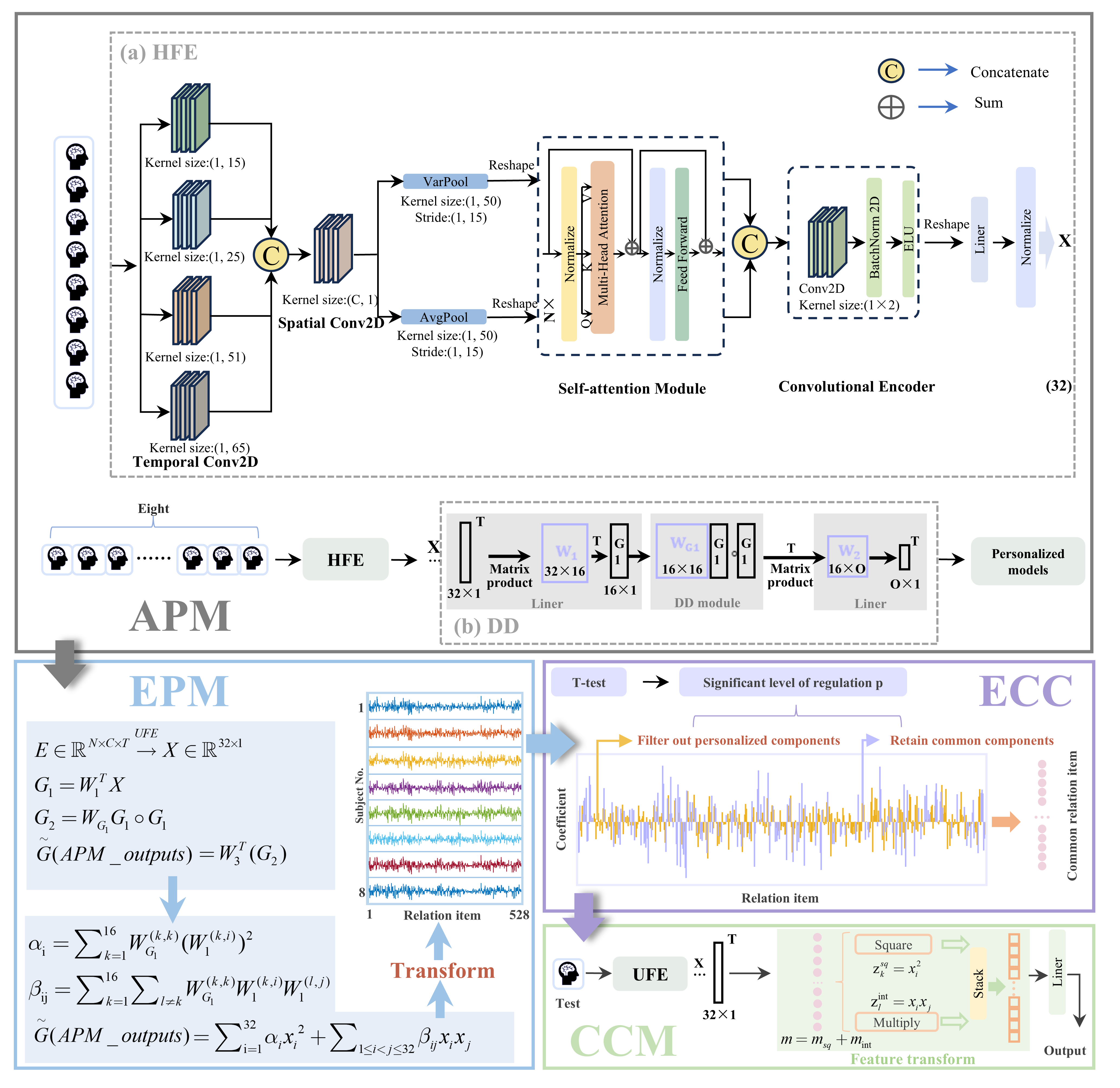}  
    \caption{Overall model architecture of the proposed CSD-SF for cross-subject EEG decoding for MI. The proposed CSD-SF consists of four major modules. 1. APM. 2. ERS. 3. ECC. 4. CCM. m: the number of common relation items. O: the number of outputs. Q: query. K: key. V: value.}
    \label{fig:li3}
    % \vspace{-10pt}  % 减少底部空白
\end{figure*} 
\subsection{Application of the System Filter to Cross-Subject MI-EEG Decoding}
To validate the effectiveness of the system filter in cross-subject EEG decoding tasks, this section applies it to MI-EEG decoding tasks, referred to as CSD-SF. The method is mainly divided into four steps. (1) Acquisition of the personalized model (APM), (2) Expansion of the personalized model (EPM), (3) Extraction of common components (ECC), (4) Construction of a cross-subject decoding model based on common components (CCM) (see Fig.~\ref{fig:li3}).
\subsubsection{Dataset}
\begin{figure}
    \centering
    \includegraphics[width=1.0\linewidth]{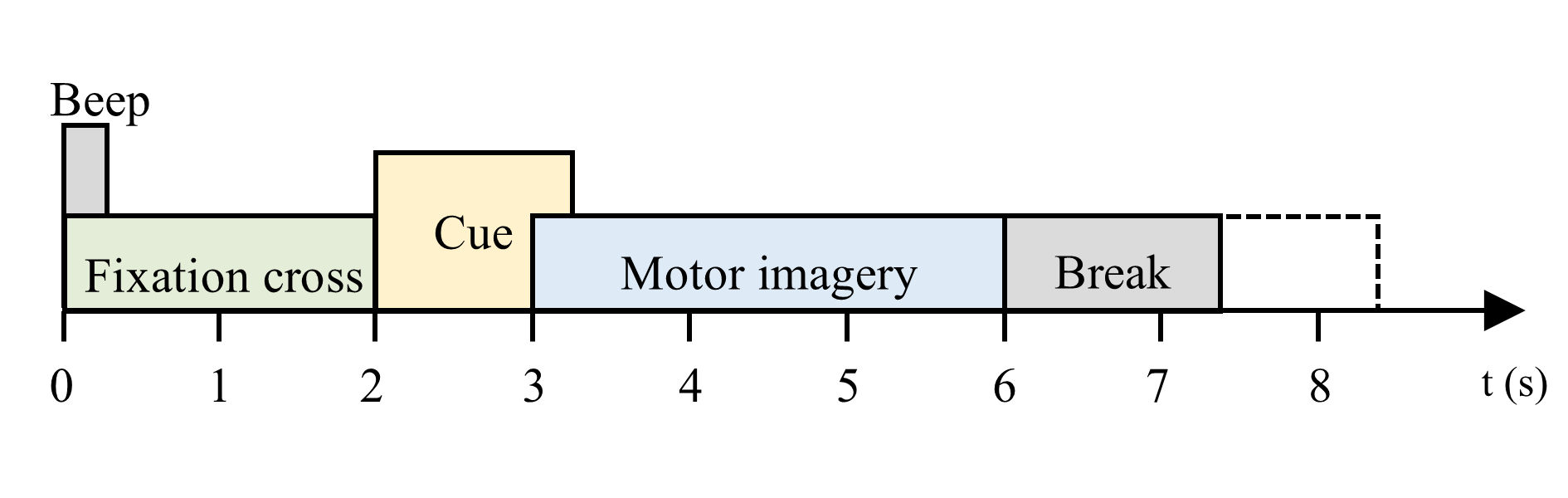}
    \caption{The experimental paradigm for BCIC IV 2a}
    \label{fig:li4}
\end{figure}

The BCIC IV 2a dataset contains EEG signals from 22 electrodes, recorded from nine healthy subjects over two sessions conducted on two different days. Each session consists of 288 four-second motor imagery trials per subject, involving the imagined movements of the left hand, right hand, feet, and tongue. This study uses data from 0 to 4 seconds. Before release, the signals were sampled at 250 Hz and bandpass filtered between 0.5 Hz and 100 Hz. The original dataset uses 288 trials from the first session (“T”) for training and 288 trials from the second session (“E”) for testing. In the cross-subject scenario, the original dataset needs to be re-split by subject using a leave-one-subject-out approach for modeling based on common components. Consequently, nine datasets (A01-A09) are obtained, each containing 576 trials (288 trials $\times$ 2 sessions) for testing one subject, with the remaining 4608 trials (288 trials $\times$ 2 sessions $\times$ 8 subjects) used for training. This procedure yields nine folds, ensuring that each subject serves as the unseen test subject once. The BCIC IV 2a dataset is publicly available as described in \cite{brunner2008}. The experimental paradigm of the BCIC IV 2a dataset is shown in Fig.~\ref{fig:li4}.

\subsubsection{Acquisition of the Personalized Model (APM)}
To obtain individualized models, hybrid features extracted by the hybrid feature extraction (HFE) framework were fed into the dendrite network (DD), a model proposed by our research group\cite{liu2021dendrite,gangmay,liu2022relation}, for training. As a white-box model with a linear–Hadamard structure, DD naturally corresponds to a finite-order polynomial expansion, providing an explicit representation that enables subsequent transformation into relation spectrums for filtering.

\textbf{HFE}: The HFE module is implemented using an end-to-end convolutional neural network adapted from Ma et al.\cite{ma2024attention} . EEG data from eight subjects are used to train a universal model. Raw EEG signals are denoted as $E \in \mathbb{R}^{N \times C \times T}$, where $N$ is the number of trials, $C$ is the number of electrodes, and $T$ indicates the number of sample points. Temporal convolutional layers with kernel sizes $(1,15), (1,25), (1,51), (1,65)$ extract multi-scale temporal features, which are subsequently processed by spatial convolution across channels $(C,1)$ to capture inter-channel dependencies. The outputs are further refined by varpool and avgpool layers, and integrated using a multi-head self-attention mechanism. Finally, a 32-dimensional normalized feature vector is obtained via linear dimensionality reduction and used as the input to the DD model.
\begin{equation}
    \boldsymbol{X} \in \mathbb{R}^{32 \times 1}
\end{equation}

\textbf{DD}: First, a dimensionality reduction module projects the input vector $\boldsymbol{X} \in \mathbb{R}^{32 \times 1}$ into a 16-dimensional subspace to enhance computational efficiency while preserving essential information.
\begin{equation}
    \boldsymbol{G}_1 = \boldsymbol{W}_1^T \boldsymbol{X}
    \label{eq1}
\end{equation}
where $\boldsymbol{G}_1 \in \mathbb{R}^{16 \times 1}$ expresses the outputs of the dimensionality reduction module. $\boldsymbol{W}_1 \in \mathbb{R}^{32 \times 16}$ is the weight matrix. $\boldsymbol{X} \in \mathbb{R}^{32 \times 1}$ expresses the inputs of the dimensionality reduction module and the outputs of the UFE module. The DD module in this paper is expressed as follows.
\begin{equation}
    \boldsymbol{G}_2 = \boldsymbol{W}_{G_1} \boldsymbol{G}_1 \circ \boldsymbol{G}_1
    \label{eq2}
\end{equation}
where $\boldsymbol{G}_1 \in \mathbb{R}^{16 \times 1}$ and $\boldsymbol{G}_2 \in \mathbb{R}^{16 \times 1}$ express the inputs and outputs of the DD module, respectively. $\boldsymbol{W}_{G_1} \in \mathbb{R}^{16 \times 16}$ is the weight matrix. $\circ$ expresses Hadamard product. Finaly, the linear module, whose dimensions are determined by the number of output categories.
\begin{equation}
    \boldsymbol{G}^{\text{APM\_outputs}} = \boldsymbol{W}_2^T \boldsymbol{G}_2
    \label{eq3}
\end{equation}
where $\boldsymbol{G}_2 \in \mathbb{R}^{16 \times 1}$ express the outputs of the DD module. $\boldsymbol{W}_2 \in \mathbb{R}^{16 \times O}$ is the weight matrix. $\boldsymbol{G}^{\text{APM\_outputs}} \in \mathbb{R}^{O \times 1}$ is the outputs of APM. $O$ is the number of categories determines. 

\subsubsection{Expansion of the Personalized Model (EPM)-Expansion}
To generate individualized relation spectrums for multi-subject analysis, the linear classification module is modified by replacing the class-specific weight matrix $\boldsymbol{W}_2 \in \mathbb{R}^{16 \times O}$ with a subject-specific vector $\boldsymbol{W}_3 \in \mathbb{R}^{16 \times 1}$, where all elements are set to 1. The revised model is expressed as follows.
\begin{equation}
    \left\{
    \begin{array}{l}
        \boldsymbol{G}_1 = \boldsymbol{W}_1^T \boldsymbol{X}\\[8pt]
        \boldsymbol{G}_2 = \boldsymbol{W}_{G_1} \boldsymbol{G}_1 \circ \boldsymbol{G}_1\\[8pt]
        \widetilde{G}^{\text{APM\_outputs}} = \boldsymbol{W}_3^T (\boldsymbol{W}_{G_1} \boldsymbol{G}_1 \circ \boldsymbol{G}_1)
    \end{array}
    \right.
    \label{eq4}
\end{equation}
where $\widetilde{G}^{APMoutputs} \in \mathbb{R}^{1 \times 1}$ represents the revised outputs of APM, which are used for transforming relation spectrums. 

The expanded output can be expressed as follows.  
\begin{equation}
    \left\{
    \begin{array}{l}
        \alpha_i = \sum_{k=1}^{16} \boldsymbol{W}_{G_1}^{(k,k)} \left( \boldsymbol{W}_1^{(k,i)} \right)^2\\[8pt]
        \beta_{ij} = \sum_{k=1}^{16} \sum_{l \neq k} \boldsymbol{W}_{G_1}^{(k,l)} \, \boldsymbol{W}_1^{(k,i)} \, \boldsymbol{W}_1^{(l,j)}\\[8pt]
        \widetilde{G}^{\text{APM\_outputs}} = \sum_{i=1}^{32} \alpha_i x_i^2 + \sum_{1 \leq i < j \leq 32} \beta_{ij} x_i x_j
    \end{array}
    \right.
    \label{eq5}
\end{equation}

\textbf{Controlling the Number of Common Components}: To ensure interpretability and computational feasibility, the number of common components must be rigorously controlled. This is achieved by leveraging the inherent completeness property of the relation spectrum. Specifically, its construction explicitly preserves all quadratic interactions in the feature space. For a d-dimensional input vector $\boldsymbol{X} \in \mathbb{R}^{d \times 1}$, the relation spectrum comprises $d$ squared items ($x_i^2$) and $\frac{d(d-1)}{2}$ pairwise interaction items ($x_i x_j$), resulting in a deterministic theoretical term count.
\begin{equation}
    N_t = d + \frac{d(d-1)}{2} = 528 \quad (d = 32).
\label{eq:eq6}
\end{equation}

This ensures that, for a 32-dimensional feature vector, the relation spectrum contains exactly 528 distinct terms, encompassing all possible squared and pairwise interaction components. 

\subsubsection{Extraction of Common Components (ECC)—Filtering}
Common components are extracted from the relation spectrums of eight subjects through statistical testing. For each quadratic term $\alpha_i x_i^2$ or $\beta_{ij} x_i x_j$ interaction term in the spectrum, coefficient values $\left\{ \alpha_i^{(1)}, \alpha_i^{(2)}, \ldots, \alpha_i^{(8)} \right\}$ or $\left\{ \beta_{ij}^{(1)}, \beta_{ij}^{(2)}, \ldots, \beta_{ij}^{(8)} \right\}$ are collected across subjects. 

A one-sample $t$-test is applied to determine whether the mean coefficient significantly deviates from zero, thereby implementing the filtering operation of the system filter. The definitions of the common and personalized components are summarized in Table~\ref{tab:feature_definitions}. The null and alternative hypotheses are defined as follows.
\begin{equation}
H_0: \mu_{R_i} = 0, \quad H_1: \mu_{R_i} \neq 0
\end{equation}
where $R_i$ denotes the set of coefficients corresponding to the $i$-th relation item across different subjects. 

The test statistic is calculated as follows.
\begin{equation}
t = \frac{\overline{R_i}}{s_i / \sqrt{N_s - 1}}
\end{equation}
where $\overline{R_i}$ is the sample mean, $s_i$ is the sample standard deviation, and $N$ is the number of subjects. A significance threshold $\alpha_{\text{threshold}}$ is set according to the experimental requirements. Items with $p$-values below the threshold were identified as common components, which were retained as the target components, whereas items failing to reach significance were classified as personalized components and filtered out as unnecessary.
\begin{table}[ht]
\caption{Definitions of common components and personalized components}
\label{tab:feature_definitions}
\centering
\begin{tabular}{@{}p{2.5cm}p{6cm}@{}}
\toprule
Feature type & Definition \\
\midrule
common components & Relation items that are consistent across subjects, where the relationships are similar (mostly positive or negative). The accumulated average significantly differs from zero, and the random component is non-zero. \\  
\addlinespace
personalized components & Relation items that are specific to each subject, with relationships that are random (can be positive or negative). The accumulated average approaches zero, and the random component is zero. \\
\bottomrule
\end{tabular}
\end{table}

\subsubsection{Construction of a Cross-subject Decoding Model based on Common Components (CCM)—Reconstruction}
The retained common components are used to reconstruct a cross-subject decoding model, which was subsequently applied to unseen subjects to evaluate decoding performance. To map all subjects’ information into a unified feature space, a feature transformation layer is designed. This layer transforms the input brain signal $\boldsymbol{X} \in \mathbb{R}^{d \times 1}$ into a new feature vector $\boldsymbol{Z} \in \mathbb{R}^{m \times 1}$ , where $m \leq N_t$ and $N_t = d + \frac{d(d-1)}{2}$ is the theoretical maximum relation items. 

The transformation involves two types of operations: squared items and interaction items. For each significant squared item $x_i^2$ , it can be described as follows.
\begin{equation}
z_k^{\text{sq}} = x_i^2
\end{equation}
where $i$ belongs to the set of indices $
S_{\text{sq}}$  corresponding to statistically validated squared features.
Similarly, for each significant interaction term $x_i x_j$ , it is represented as follows.
\begin{equation}
z_l^{\text{int}} = x_i x_j
\end{equation}
where the pair $(i, j)$ is included in the set $S_{\text{int}}$ of significant interactions. The final transformed feature vector $\boldsymbol{Z}$ is constructed by concatenating all retained components.
\begin{equation}
\boldsymbol{Z} = \left[ z_1^{\text{sq}}, \dots, z_{m_{\text{sq}}}^{\text{sq}}, z_1^{\text{int}}, \dots, z_{m_{\text{int}}}^{\text{int}} \right]^\top \in \mathbb{R}^{m \times 1}
\end{equation}
where $m = m_{\text{sq}} + m_{\text{int}}
$ , ensuring that only cross-subject common features are preserved. The linear classifier then maps the m-dimensional feature vector $\boldsymbol{Z}$ to class probabilities through a linear transformation followed by softmax activation. The linear transformation is defined as follows.
\begin{equation}
% h = \mathbf{W}^\top \mathbf{Z} + b
\boldsymbol{h} = \boldsymbol{W}^\top \boldsymbol{Z} + \boldsymbol{b}
\end{equation}
where $\boldsymbol{W} \in \mathbb{R}^{m \times O}$ is the weight matrix, $\boldsymbol{b} \in \mathbb{R}^{O \times 1}$ is the bias vector, and $O$ is the number of categories. Finally, the output probabilities are computed via the softmax function. The model is trained by minimizing the cross-entropy loss. 

\section{Experiments}
\subsection{Experimental Details}
\subsubsection{Performance Metric}
In the experiment, classification accuracy and Cohen's kappa coefficient are employed to evaluate the decoding performance of different networks, thereby comprehensively assessing the performance of the model in cross-subject tasks. The method for calculating accuracy is described as follows.
\begin{equation}
    \text{Accuracy} = \frac{TP + TN}{TP + TN + FP + FN}
\end{equation}
where $TP$ and $TN$ represent the number of correct positive and correct negative samples predicted by the model, respectively, while $FP$ and $FN$ denote the number of false positive and false negative samples predicted by the model, respectively. 

The method for calculating Cohen's kappa coefficient is described as follows.
\begin{equation}
    k = \frac{P_o - P_e}{1 - P_e}
\end{equation}
where $P_o$ denotes the accuracy of the model, and $P_e$ represents the probability or accuracy of a random guess.

We evaluate the statistical significance between the proposed model and other baseline models using paired $t$-tests and $p$-values.

\subsubsection{Evaluation Baselines}
We compare the proposed CSD-SF with six baseline models.

ShallowConvNet \cite{shallowconvnet} consists of two convolutional layers, which are used for temporal and spatial filtering, respectively.

EEGNet \cite{EEGNet} is a lightweight neural network that first learns through temporal filtering and then performs global spatial filtering using deep convolutions. After extracting initial temporal and spatial features, it utilizes separable convolutions to perform deep temporal feature extraction, demonstrating a reduced parameter count. 

CRAM \cite{CRAM} is a graph-based convolutional recurrent attention model, which first introduces a graph structure to represent the EEG node location information. Subsequently, the convolutional recurrent attention model learns EEG features from both spatial and temporal dimensions.

CCNN \cite{CCNN} is a multilayer CNN approach that integrates CNNs with different characteristics and structures, utilizing convolutional features from different layers to capture spatial and temporal features from raw EEG data. 

A novel EEG 3D representation is introduced in the multi-branch 3D CNN \cite{3D-CNN} , which preserves both spatial and temporal information. Based on this 3D representation, a multi-branch 3D CNN is used to extract MI-related features.

CTNet \cite{ctnet} is a model for EEG-based MI classification that combines convolutional and Transformer architectures. It first extracts local and spatial features using a convolutional module, then captures global dependencies with a Transformer encoder, and finally classifies the signals using fully connected layers.
\begin{figure*}[htbp]
    \centering
    \includegraphics[width=1.0\textwidth]{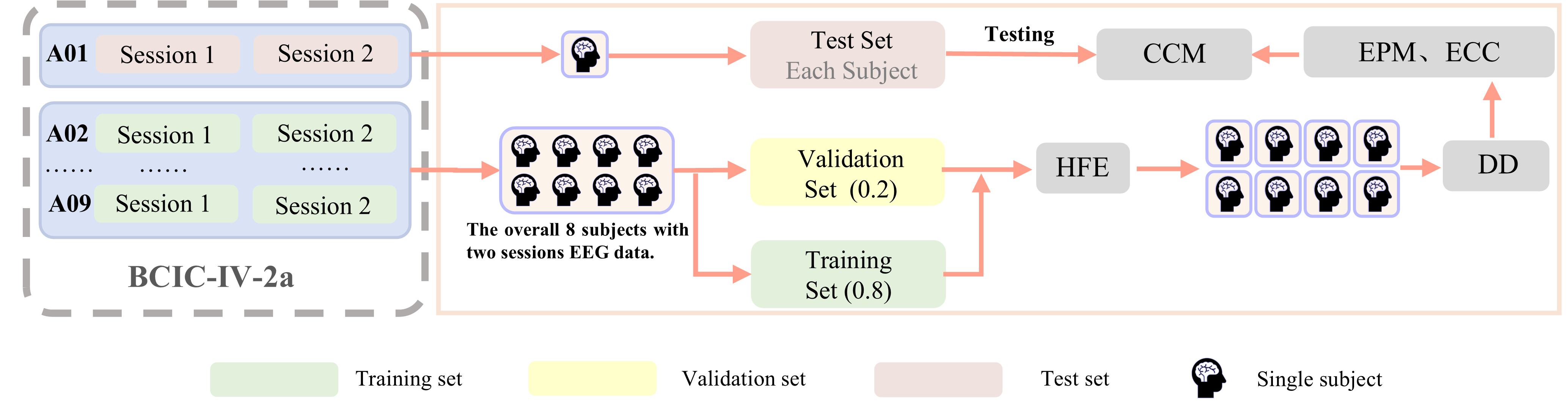}  
    \caption{Cross-subject experimental setting.}
    \label{fig:li5}
\end{figure*}

\subsection{Experimental Setups}
All experiments in this paper are conducted on a single GPU, specifically the Nvidia RTX 3080, using PyTorch, while the extraction of common relation items is performed using MATLAB. The cross-subject setting for the entire experiment is illustrated in Fig.~\ref{fig:li5}.

For the upcoming experiments, this paper adopts the Leave-One-Subject-Out (LOSO) validation strategy. This method is widely regarded as the most rigorous approach for cross-subject classification, as it completely excludes certain subjects during training and evaluates the model's performance on unseen subjects\cite{EEGNet,114,44,54}. It provides an objective measure of generalization ability. 

To avoid ambiguity, the Table~\ref{tab:notation_definitions} specifies the notational conventions used for different subject configurations.
\begin{table}[ht]
\caption{Notation used for subject configurations.}
\label{tab:notation_definitions}
\centering
\begin{tabular}{@{}p{1.5cm}p{6cm}@{}}
\toprule
Notation  & Definition \\
\midrule
subject i & Refers to subject A0i (i = 1, 2, ..., 9)  \\  
\addlinespace
Subject -i & Indicates that data from subject A0i is used for testing, while data from the remaining eight subjects is used for training (e.g., leave-one-subject-out setting) (i = 1, 2, ..., 9) (see Fig.~\ref{fig:li5}).\\
\bottomrule
\end{tabular}
\end{table}
\begin{figure*}[htbp]
    \centering
    \includegraphics[width=1.0\textwidth]{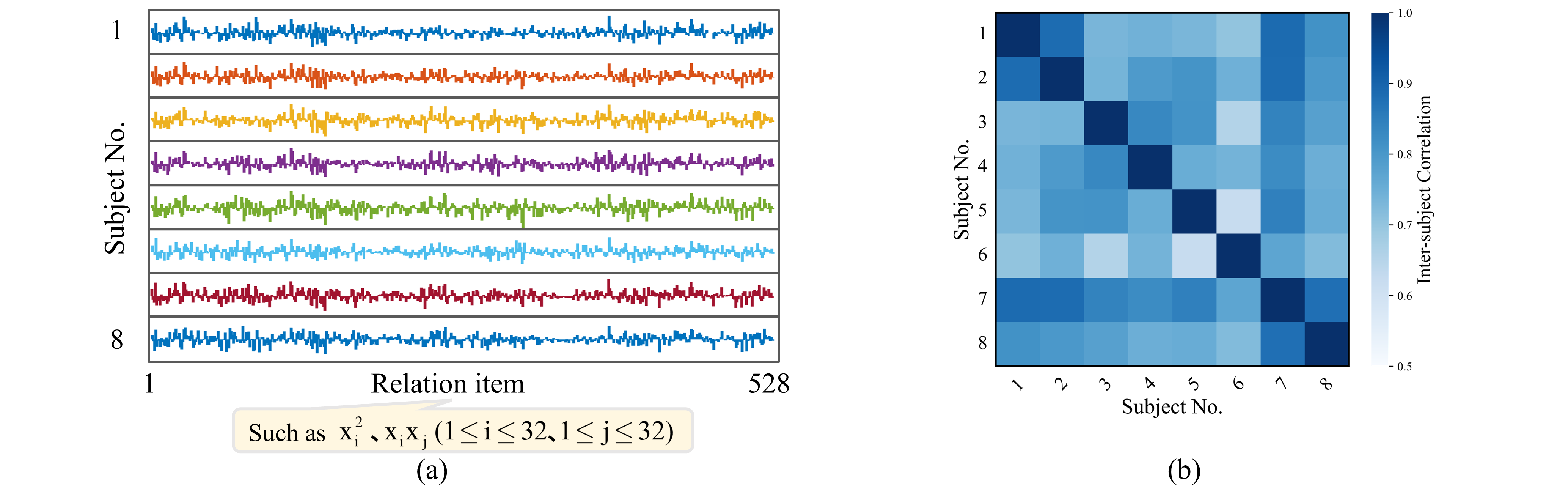}  
    \caption{Transform APM models. (Test data for subject 9 and train data for the remaining eight subjects). (a) The relation spectrum of subject 1-subject 8. (b) Correlation matrix between eight personalized models.}
    \label{fig:li6}
\end{figure*}

\subsubsection{Personalized Model Training and Acquisition}
Hybrid features obtained through HFE were used as input to the DD for training. Using the LOSO strategy, EEG data from eight subjects were split into training and validation sets at an 8:2 ratio. Five-fold cross-validation was used for iterative optimization, with a learning rate of 0.0002, batch size of 64, and 500 training epochs. For personalized modeling, the trained network was further adapted to each subject using their individual training data, resulting in eight subject-specific models. Hyperparameters were optimized via grid search, with the learning rate kept at 0.0002, batch size increased to 256, and 200 epochs of training.
Finally, the subject-specific model was applied to the test set of the left-out subject, and classification accuracy was evaluated to assess the decoding performance of the personalized models. 

\subsubsection{Extraction of Common components}
Based on the optimal parameters from model training, the eight subject-specific models were expanded into their corresponding relation spectrums. A one-sample $t$-test was performed on the coefficients of all 528 relation items to assess whether their mean values across subjects significantly differed from zero. Relation items passing the significance test were regarded as common components, with predefined thresholds of $\alpha_{\text{threshold}} = 0.5, 0.1, 0.01,$ and $0.001$ used to analyze their impact on cross-subject decoding performance.

\subsubsection{Building Cross-Subject Models with Common Components}
To evaluate the proposed CSD-SF framework under cross-subject conditions, the LOSO strategy was adopted. In each iteration, data from eight subjects were used for training, while the remaining subject was held out for testing, ensuring evaluation on unseen individuals. This process was repeated nine times, and the average decoding accuracy across subjects was reported. 

Model training in this stage was conducted with a learning rate of 0.0001, a batch size of 32, and 200 epochs. For each subject, the subset of common components yielding the highest decoding accuracy was selected. Finally, statistical comparisons between different methods were performed using paired-sample $t$-tests on the BCIC IV 2a dataset.

\begin{figure*}[htbp]
    \centering
    \includegraphics[width=1.0\textwidth]{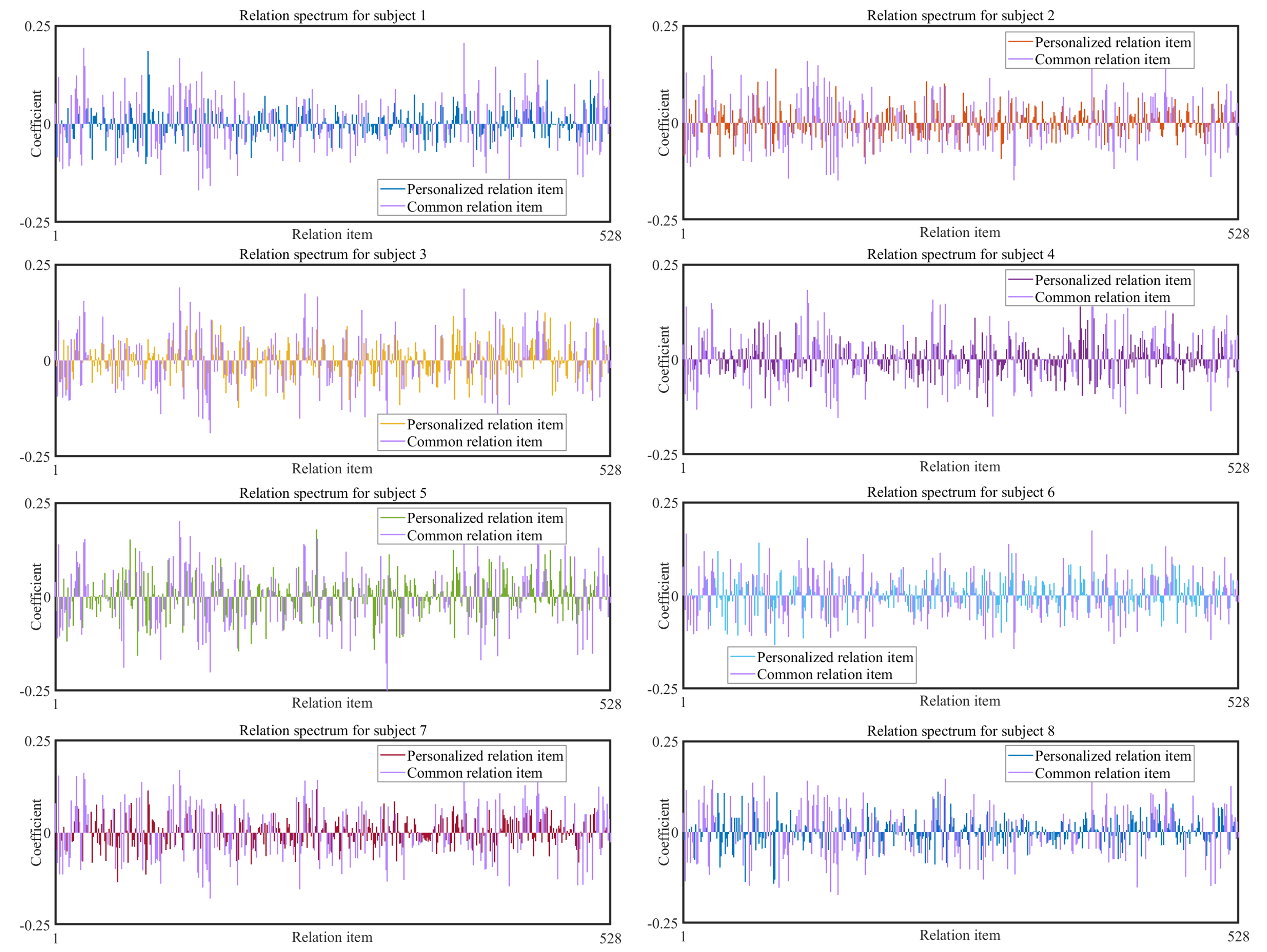}  
    \caption{Visualization of common and personalized components. (e.g. Subject-9)}
    \label{fig:li7}
\end{figure*}

\begin{table*}[htbp]  
\centering
\caption{Cross-subject classification results for BCIC IV 2a dataset. Bold denotes the best numerical values. * denotes a significant difference ($p < 0.05$), while ** denotes a highly significant difference ($p < 0.01$).}  
\resizebox{\textwidth}{!}{  
\begin{tabular}{lccccccccccccc}  
\toprule
Methods & Subject-1 & Subject-2 & Subject-3 & Subject-4 & Subject-5 & Subject-6 & Subject-7 & Subject-8 & Subject-9 & Accuracy & Standard & Kappa & $p$-value\\
\midrule
EEGNET & 53.76 & 39.54 & 54.88 & 43.02 & 51.8 & 48.96 & 60.70 & 61.38 & 47.82 & 51.32 & 7.34 & - & ** \\
ShallowConvNet & 68.92 & 43.75 & 75.17 & \textbf{55.73} & 43.75 & 45.31 & 74.65 & \textbf{80.03} & 66.84 & 61.57 & 14.63 & - & - \\
Multi-branch 3D CNN & 49.51 & 40.74 & 64.5 & 44.56 & 54.29 & 40.46 & 58.87 & 59.75 & 56.83 & 52.17 & 8.76 & 0.453 & ** \\
CCNN & 62.07 & 42.44 & 63.12 & 52.09 & 49.96 & 37.16 & 62.54 & 59.32 & \textbf{69.43} & 55.35 & 10.66 & - & ** \\
CRAM & 61.02 & 42.35 & 73.11 & 50.43 & 50.74 & 51.48 & 67.26 & 69.72 & 66.85 & 59.22 & 10.74 & - & ** \\
CTNet & \textbf{69.27} & 43.92 & 79.34 & 55.38 & 43.92 & 36.11 & 65.10 & 70.66 & 64.06 & 58.64 & 14.61 & 0.449 & * \\
CSD-SF & 69.10 & \textbf{49.31} & \textbf{80.38} & 55.03 & \textbf{63.02} & \textbf{54.17} & \textbf{76.04} & 70.83 & 65.80 & \textbf{64.85} & 10.47 & 0.531 & - \\
\bottomrule
% \label{table 1}
\end{tabular}
}
\\[1ex]
\footnotesize{*Note: Chance level for the four-class task is 25\%; despite inter-subject variability, CSD-SF consistently surpasses this baseline.}
\label{table 1}
\end{table*}

\section{Results}
\subsection{Relation Spectrum of the Personalized Model}
Fig.~\ref{fig:li6}(a) depicts the relation spectrum of the personalized models for subject 1 to subject 8, with each curve representing the relation spectrum of an individual subject. The x-axis corresponds to 528 relation items, and the y-axis represents the coefficients associated with these relation items. Upon observation, significant differences can be observed between the relation spectrums of different subjects, reflecting the influence of individual differences on EEG signal decoding. This shows that the model training process successfully captured EEG features of each subject. Furthermore, Fig.~\ref{fig:li6}(b) presents the correlation matrix of the eight personalized models. In the matrix, the intensity of the color blocks indicates the similarity of relation spectrums between subjects, with darker blocks representing higher correlation and lighter blocks representing lower correlation. It can be observed that, while there is some similarity between subjects, individual differences remain prominent. In particular, the relation spectrums of certain subjects are highly similar, highlighting the potential for cross-subject feature sharing. 

\subsection{Common Components}
\subsubsection{Visualization of Common Components}
To validate the ability of the proposed CSD-SF model to extract common components, we performed a key visualization analysis. Fig.~\ref{fig:li7} displays the extracted common components across different subjects, based on data from subject 1 to subject 8 (Subject-9), along with the relation spectrum analysis results. This visualization allows us to observe the common relationships between different subjects intuitively. These features represent stable components shared by multiple subjects, providing visual evidence to confirm whether common components with broad applicability have been successfully extracted. 

\subsubsection{The Number of Common Components Corresponding to Different Predefined Thresholds}
\begin{figure}
    \centering
    \includegraphics[width=1.0\linewidth]{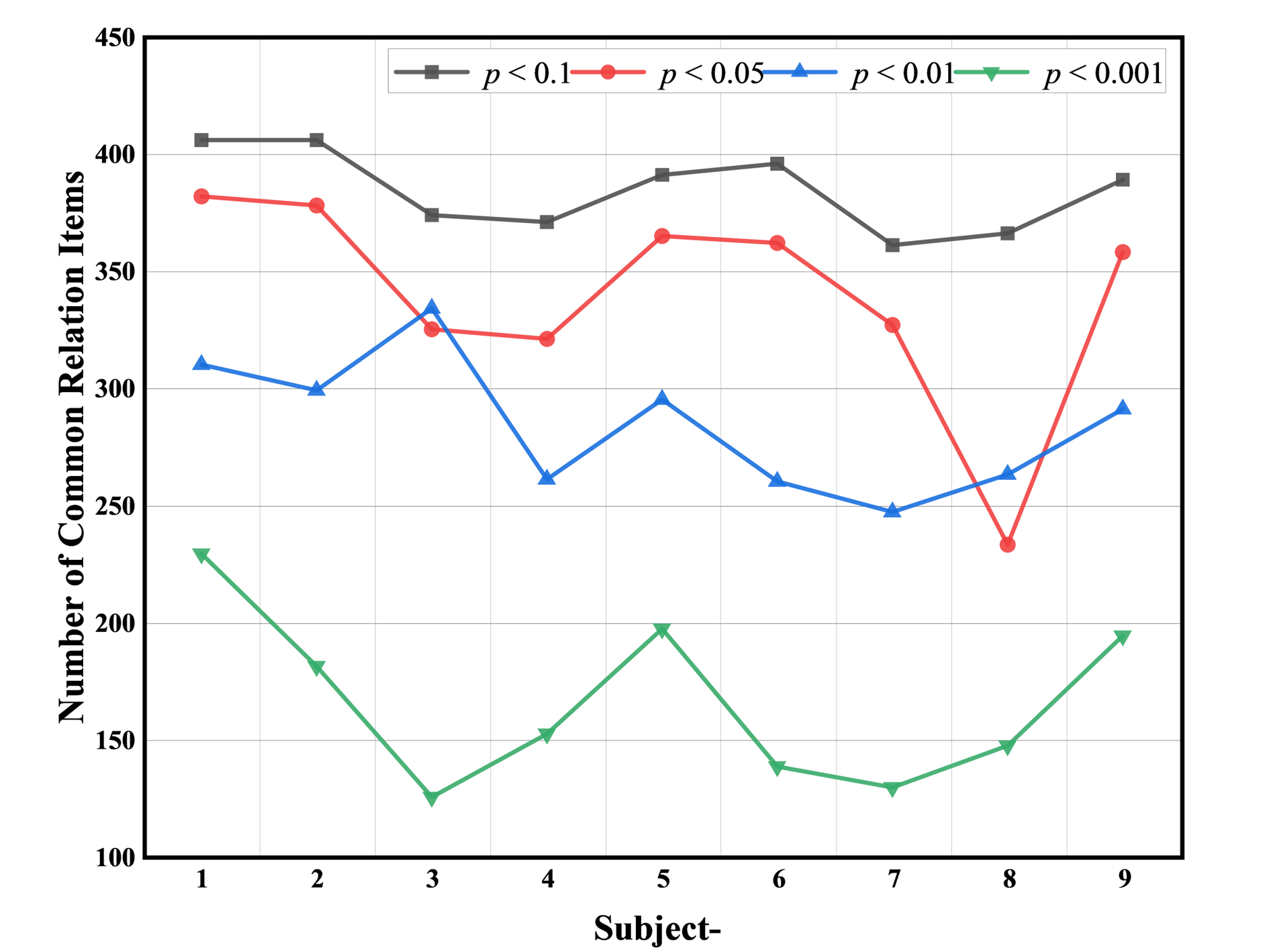}
    \caption{The change in the number of common relation items in the CSD-SF on the BCIC IV 2a dataset is observed as the predefined threshold $\alpha_{\text{threshold}}$ is gradually increased from 0.001 to 0.1.}
    \label{fig:li8}
\end{figure}

As shown in Fig.~\ref{fig:li8}, the number of common relation items decreases with stricter predefined thresholds $\alpha_{\text{threshold}}$. When $\mathit{p} < 0.1$, the number of features ranges from 361 to 406 across subjects. As the threshold is reduced to 0.05, 0.01, and 0.001, fewer features are retained, with the number of features dropping to as low as 125 for Subject-3 when $\mathit{p} < 0.001$. This indicates that stricter thresholds lead to fewer but more statistically significant components. 

\subsection{Decoding Performance Evaluation}
\subsubsection{Decoding Performance Across Different Models}
Table~\ref{table 1} comprehensively presents the performance of the proposed CSD-SF compared with other baseline models on the BCIC IV 2a dataset. 

Firstly, in terms of average classification accuracy, the proposed CSD-SF outperformed EEGNet by 13.53\%, with a statistically significant difference ($\mathit{p} < 0.01$). This demonstrates the superior robustness and generalization ability of CSD-SF in extracting common components and enabling cross-subject decoding. Additionally, compared with CTNet, CSD-SF achieved a 6.21\% higher average accuracy and a higher kappa value, while also demonstrating a lower standard deviation ($\mathit{p} < 0.05$), indicating more stable performance across subjects.
When compared with other baseline models, such as the Multi-branch 3D CNN, CCNN, and CRAM, CSD-SF achieved classification accuracy improvements of 12.68\%, 9.23\%, and 5.63\%, respectively, with these differences also being statistically significant ($\mathit{p} < 0.01$). These results further validate the superiority of CSD-SF in cross-subject tasks, particularly in its effective utilization of common components during feature extraction and model training to reduce the impact of individual differences on decoding performance. Although CSD-SF achieved a 3.28\% improvement over ShallowConvNet, the difference was not statistically significant ($\mathit{p} > 0.05$). This suggests that these models may perform similarly in some cases, which may be attributed to the limitations of these baseline models in handling inter-subject differences. This further highlights the advantages of CSD-SF in terms of adaptability and generalization.

\begin{figure}
    \centering
    \includegraphics[width=1.0\linewidth]{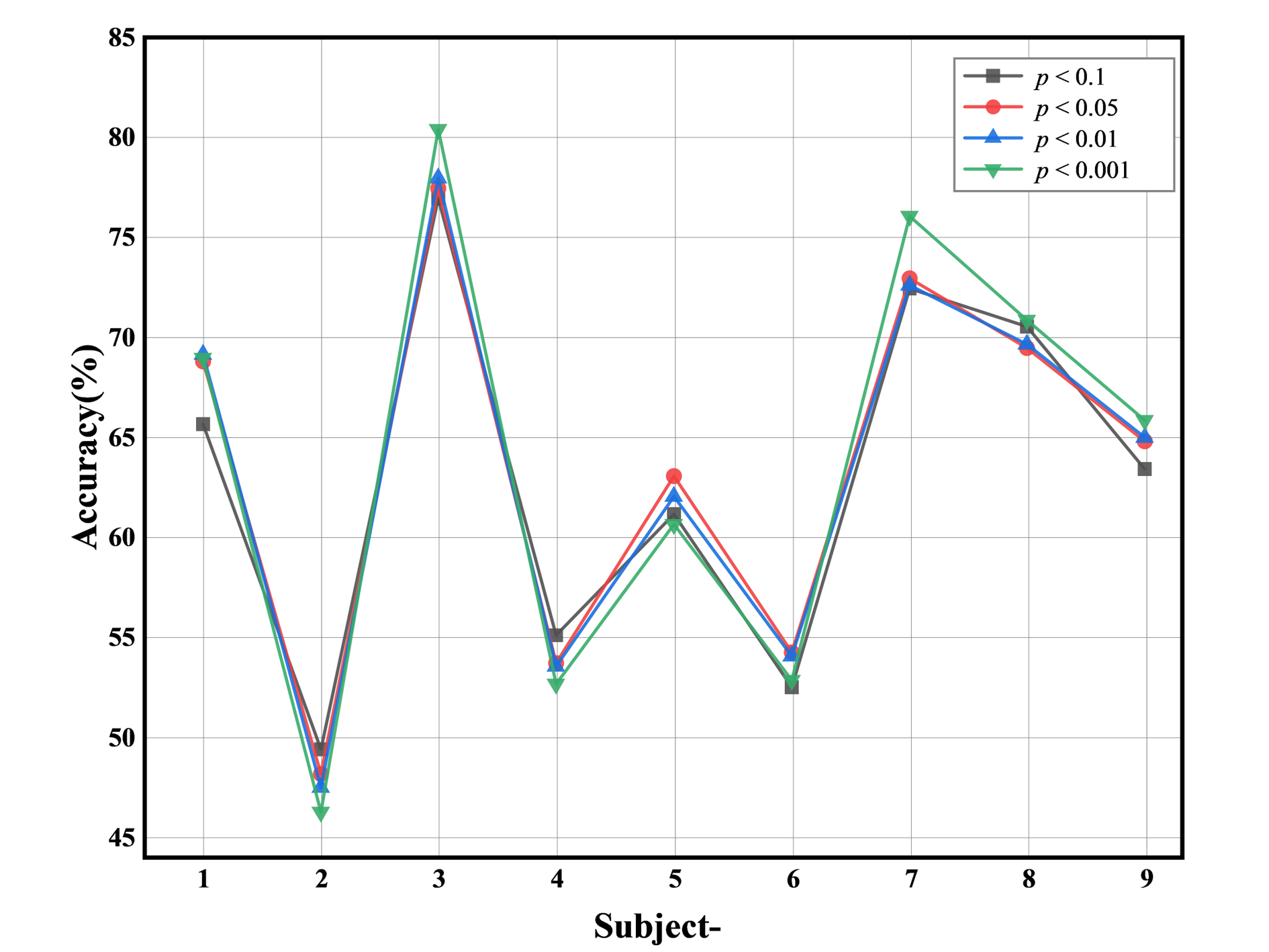}
    \caption{Cross-subject decoding results based on the BCIC IV 2a dataset, corresponding to the number of common relation items under different predefined thresholds $\alpha_{\text{threshold}}$.}
    \label{fig:li9}
\end{figure}

\subsubsection{Effect of Common Components Extraction on Cross-Subject Decoding Performance}
The classification accuracy results based on these thresholds are summarized in Fig.~\ref{fig:li9}. When $\mathit{p} < 0.1$, the average accuracy was 62.96\%, with subject-specific accuracies ranging from 49.31\% (Subject-2) to 65.62\% (Subject-1). Performance improved as the $\mathit{p}$-value threshold was reduced. When $\mathit{p} < 0.05$, accuracy increased to 63.58\%, and when $\mathit{p} < 0.01$, it rose to 64.93\%, with Subject-4, Subject-6, and Subject-8 showing substantial improvement. However, when $\mathit{p} < 0.001$, accuracy reached 63.79\%, showing a marginal improvement over when $\mathit{p} < 0.01$, indicating that further reduction in features may not provide significant performance gains.

Paired sample $t$-test was conducted to assess the statistical significance of the performance differences across different predefined thresholds. Results showed significant improvements when $\mathit{p} < 0.05$ and $\mathit{p} < 0.01$, indicating that stricter feature selection enhances the model’s generalization ability. However, the difference between when $\mathit{p} < 0.01$ and when $\mathit{p} < 0.001$ was not statistically significant, suggesting diminishing returns at very strict thresholds.
\begin{figure}
    \centering
    \includegraphics[width=1.0\linewidth]{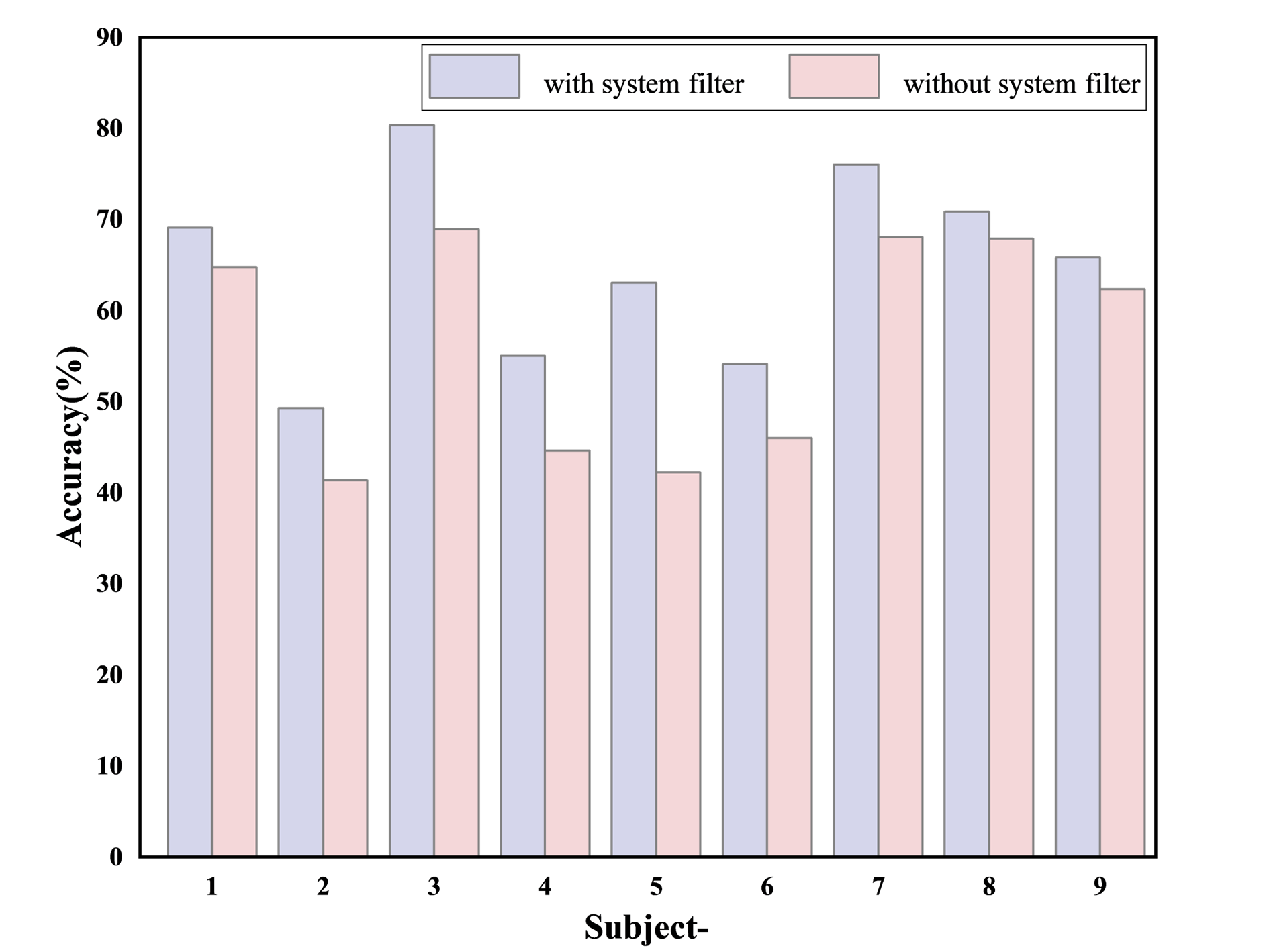}
    \caption{Cross-subject decoding accuracy with and without the system filter.}
    \label{fig:li10}
\end{figure}

\begin{figure*}[htbp]
    \centering
    \includegraphics[width=1.0\textwidth]{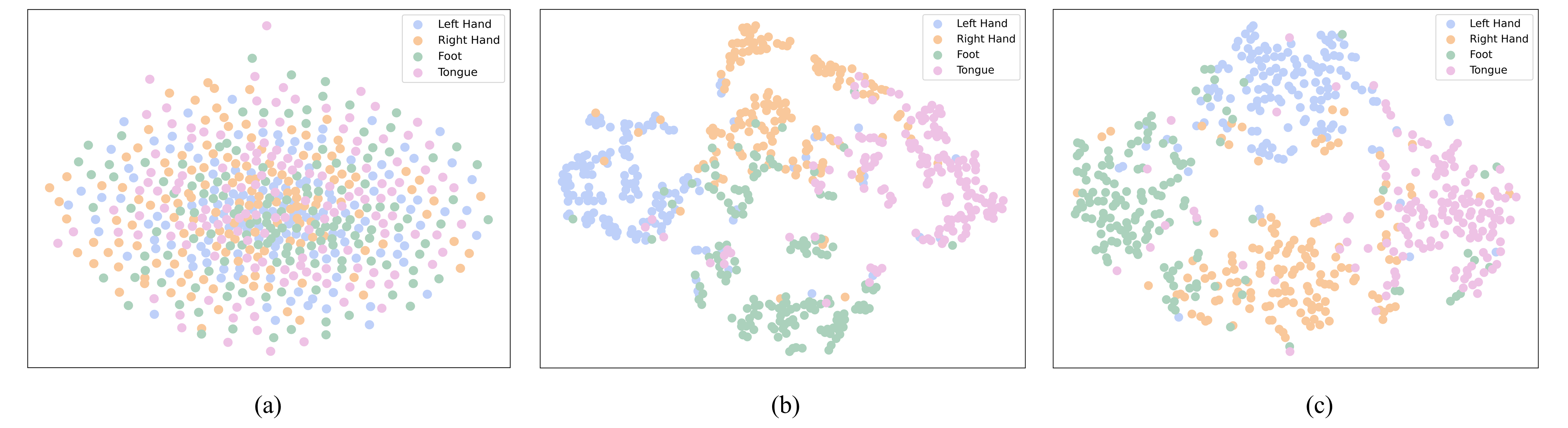}
    \caption{The visualization for Subject-3. (a) T-SNE visualization of raw data. (b) T-SNE visualization of data without using the system filter. (c) T-SNE visualization of data using the system filter (CSD-SF). Different colors represent different categories.}
    \label{fig:li11}
\end{figure*}
\subsection{Ablation Study}
To evaluate the effectiveness of the proposed system filter in the CSD-SF model for improving cross-subject EEG decoding, an ablation study was performed. Specifically, we compared two conditions for cross-subject decoding using the LOSO cross-validation: with system filter (CSD-SF) and without system filter. In the “without system filter” condition, no relation spectrum transformation or filtering was applied. Instead, cross-subject decoding was performed directly using deep feature extraction followed by a fully connected layer. The experimental results are illustrated in Fig.~\ref{fig:li10}. Our proposed method achieved high classification accuracy across all subject, with significant improvements observed for Subject 3 and Subject 5. On average, the classification accuracy increased by 8.62\% compared to the “without system filter” condition  ($\mathit{p} < 0.05$). These results demonstrate that system filter achieved superior performance in the four-classification MI task, underscoring its effectiveness in enhancing cross-subject generalization.

The t-SNE(t-distributed Stochastic Neighbor Embedding)\cite{t-sne} visualizations for Subject 3 are shown in Fig.~\ref{fig:li11}, highlighting the impact of the system filter MI task separation. The raw EEG data exhibits significant overlap between the different motor imagery tasks (Left Hand, Right Hand, Foot, and Tongue), suggesting that individual-specific noise impedes the separation of tasks and complicates classification. In the condition without the system filter, the leave-one-subject-out cross-validation was performed on the extracted features without applying any transformation or component filtering. While some task separation is observed, the clusters still overlap significantly, indicating that subject-specific noise continues to affect model performance. In contrast, after applying the system filter (CSD-SF), the separation between the MI tasks becomes much clearer. By focusing on common components shared across subjects and filtering out personalized components, the system filter enhances the clarity of the clusters, improving task separation and decoding accuracy. These results demonstrate that the system filter significantly improves the cross-subject decoding performance by extracting stable, generalized features and reducing the impact of individual differences.

\section{Discussion}
In this paper, we propose a system filter that extends the traditional concept of signal filtering to the system modeling level and apply it to cross-subject MI-EEG decoding, termed CSD-SF. The system filter enables explicit decomposition of system components, filtering out personalized components and retaining common components across subjects to construct a universal decoding model. Experimental results demonstrate that CSD-SF achieves superior performance in the four-class MI task, confirming its effectiveness in improving model generalization.

This discussion highlights two key aspects: the methodological value of the system filter as a general modeling framework, and the role of the CSD-SF in developing universal cross-subject decoding models based on common components.

\subsection{System Filter as a General Modeling Framework}
The proposed system filter extends the notion of filtering from the signal domain to the system domain. Traditional signal filtering, such as Fourier or wavelet filtering, aims to remove noise while preserving the essential components of the signal. The system filter extends this concept by operating directly on the system representation, focusing on the intrinsic relationships among variables rather than the raw signal itself. By expanding the system into a polynomial representation and mapping it into the spectral domain, it enables explicit decomposition of relational components, allowing unnecessary items to be selectively filtered out.

This structural filtering provides a controllable and interpretable approach to system modeling. Rather than relying on implicit parameter adaptation or end-to-end optimization, it imposes a structural stability constraint to preserve essential mappings while filtering out unstable relations\cite{5.1-1-1,5.1-1-2,5.1-2-1,5.1-2-2,5.1-3-1}. By explicitly constraining system structure and selectively removing unwanted components, the system filter achieves interpretable and generalizable modeling. Although this paper focuses on EEG decoding, the same principle can be extended to eliminate unwanted components in other complex systems, offering a general paradigm for stable and interpretable modeling.

\subsection{Toward Universal Cross-Subject BCI Decoding Models}
Traditional approaches, including temporal–spatial filter optimization \cite{5.2-1-1,5.2-1-2,5.2-1-3} and deep learning models trained on large datasets \cite{5.2-2-1,5.2-2-2,5.2-2-3}, achieve adaptability by reinforcing cross-subject common features or accommodating subject-specific variations. However, they do not explicitly distinguish common components from personalized ones, and residual individual interference often limits their generalization ability.

In contrast, the proposed CSD-SF framework, built upon the system filter, provides an explicit and interpretable mechanism for isolating common and personalized components in cross-subject EEG decoding. The method expands individual system models into relation spectrums and applies one-sample $t$-test to evaluate the stability of each relation item. Unstable items are regarded as personalized components reflecting individual variability, whereas stable relation items consistent across subjects are retained to construct the decoding model. These stable relation items are  assumed to represent shared cortical activation patterns underlying motor imagery.

By explicitly distinguishing common from personalized relation items at the system level, CSD-SF suppresses individual noise while preserving common neural dynamics. As shown in Fig.~\ref{fig:li11}(c), the common model derived from stable relation items achieves clear task-specific separation when decoding new subjects, confirming the effectiveness of the proposed method.

Beyond the methodological contributions, the CSD-SF framework enhances the practical usability of EEG-based BCI systems. By identifying neural patterns that remain stable across individuals, it reduces the need for extensive subject-specific calibration and enables faster adaptation in clinical and rehabilitation contexts. This advantage is critical for motor-impaired users, for whom minimizing recalibration demands is essential to ensure consistent and reliable BCI performance.

\section{Conclusion}
This paper proposes a novel system filter, a method that transforms a complex system into a series of polynomial terms. By expanding the system into a spectral representation, the filter selectively removes unnecessary components, retaining only the target components necessary for the task.

In the context of cross-subject decoding, we applied the system filter within the CSD-SF framework, specifically developed for MI-EEG decoding. The personalized models were transformed into relation spectrums, and statistical testing across subjects enabled the separation of personalized components from common components. The retained common components were used to construct a universal model.

Experimental results on the BCIC IV 2a dataset demonstrate the effectiveness of the system filter in improving cross-subject decoding performance. The proposed method achieved an average accuracy improvement of 3.28\% over baseline approaches, underscoring the significance of focusing on common components to enhance cross-subject generalization. This improvement confirms that the system filter enhances the robustness of the decoding model by isolating common components and removing personalized noise, thereby improving cross-subject generalization. This is crucial for enabling effective BCI applications across diverse subjects.

\end{document}